**Demystifying Application Programming Interfaces (APIs): Unlocking the Power of Large Language Models and Other Web-based AI Services in Social Work Research**

Brian E. Perron[1], Hui Luan[2], Zia Qi[1], Bryan G. Victor[3], Kavin Goyal[4]

[1]School of Social Work, University of Michigan

[2]School of Social Development, Department of Social Work, Tianjin University of Technology

[3]School of Social Work, Wayne State University

[4] Department of Electrical Engineering and Computer Science, University of Michigan

**Author Note**

Brian E. Perron https://orcid.org/0009-0008-4865-451X

Hui Luan https://orcid.org/0000-0002-7033-9240

Bryan G. Victor https://orcid.org/0000-0002-2092-912X

# Abstract

Application Programming Interfaces (APIs) are essential tools for social work researchers aiming to harness advanced technologies like Large Language Models (LLMs) and other AI services. This paper demystifies APIs and illustrates how they can enhance research methodologies. It provides an overview of API functionality and integration into research workflows, addressing common barriers for those without programming experience. The paper offers a technical breakdown of code and procedures for using APIs, focusing on connecting to LLMs and leveraging them to facilitate API connections. Practical code examples demonstrate how LLMs can generate API code for accessing specialized services, such as extracting data from unstructured text. Emphasizing data security, privacy considerations, and ethical concerns, the paper highlights the importance of careful data handling when using APIs. By equipping researchers with these tools and knowledge, the paper aims to expand the impact of social work research through the effective incorporation of AI technologies.

Keywords: Large language models, Artificial intelligence, Application Programming Interface

**Demystifying Application Programming Interfaces (APIs): Unlocking the Power of Large Language Models and Other Web-based AI Services in Social Work Research**

The field of social work research is at a pivotal moment, experiencing a transformative shift in how data is managed and analyzed. With the advent of advanced artificial intelligence (AI) technologies—particularly Large Language Models (LLMs)—new opportunities are available for enhancing research methodologies and streamlining complex workflows. Application programming interfaces (APIs), which facilitate local and external linkages between software applications, are at the heart of this transformation, functioning as essential tools for accessing and utilizing sophisticated AI technologies. While APIs have long been integral to software development, their significance has surged with the rise of AI and LLMs. For social work researchers, APIs now act as gateways to robust AI services and vast data resources, enabling inquiries and analyses that were once impractical or impossible.

Integrating APIs into social work research unlocks unprecedented possibilities for data management and analysis. Researchers can automate tedious data processing tasks, perform large-scale analyses, and access cutting-edge AI tools—all of which can lead to more insightful findings and impactful interventions. However, leveraging APIs typically involves writing computer code, often in languages like Python, which can be daunting for social work researchers who may not have a programming background. Fortunately, recent technology advancements have dramatically lowered these barriers. The emergence of LLMs has revolutionized coding assistance by offering interactive guidance, code generation, and error correction. Moreover, user-friendly online development environments have made writing and testing API integrations more accessible. Combined with Python's intuitive syntax, these developments have democratized access to powerful data processing tools, allowing social work researchers to harness APIs regardless of their prior programming experience.

**Demystifying Application Programming Interfaces (APIs): Unlocking the Power of Large Language Models and Other Web-based AI Services in Social Work Research**

The field of social work research is at a pivotal moment, experiencing a transformative shift in how data is managed and analyzed. With the advent of advanced artificial intelligence (AI) technologies—particularly Large Language Models (LLMs)—new opportunities are available for enhancing research methodologies and streamlining complex workflows. Application programming interfaces (APIs), which facilitate local and external linkages between software applications, are at the heart of this transformation, functioning as essential tools for accessing and utilizing sophisticated AI technologies. While APIs have long been integral to software development, their significance has surged with the rise of AI and LLMs. For social work researchers, APIs now act as gateways to robust AI services and vast data resources, enabling inquiries and analyses that were once impractical or impossible.

Integrating APIs into social work research unlocks unprecedented possibilities for data management and analysis. Researchers can automate tedious data processing tasks, perform large-scale analyses, and access cutting-edge AI tools—all of which can lead to more insightful findings and impactful interventions. However, leveraging APIs typically involves writing computer code, often in languages like Python, which can be daunting for social work researchers who may not have a programming background. Fortunately, recent technology advancements have dramatically lowered these barriers. The emergence of LLMs has revolutionized coding assistance by offering interactive guidance, code generation, and error correction. Moreover, user-friendly online development environments have made writing and testing API integrations more accessible. Combined with Python's intuitive syntax, these developments have democratized access to powerful data processing tools, allowing social work researchers to harness APIs regardless of their prior programming experience.

This article aims to empower social work researchers by:

1. Demystifying APIs: We provide an accessible overview of APIs, explaining their functions and illustrating how they can be integrated into research workflows to enhance efficiency and expand capabilities.
2. Offering technical guidance: We provide a detailed breakdown of the code and procedures necessary for using APIs, focusing on connecting to LLMs. We also provide practical strategies for leveraging LLMs to facilitate API connections.
3. Presenting practical applications: Through concrete code examples, we demonstrate how LLMs can generate API code for accessing specialized services. These examples serve as templates that researchers can adapt to their specific needs, bridging the gap between theoretical understanding and practical implementation.

By addressing these objectives, we aim to equip social work researchers with the knowledge and tools to effectively incorporate AI technologies into their data analysis processes. Embracing APIs and AI tools is both a technical enhancement and a strategic advancement for social work research. By expanding the scope and depth of research, we can generate more nuanced insights, inform better policies, and ultimately improve social services. This integration positions social work researchers at the forefront of innovation, enabling them to contribute significantly to addressing complex social issues in an increasingly data-driven world.

Various code examples, computer functions, and prompts for LLMs are provided throughout the paper. These are highlighted in green to distinguish them from the main text for clarity and ease of reference.

**Primer on APIs**

**Functionality of APIs**

An API is a set of rules and tools that enables different software applications to communicate with each other. We offer the commonly used analogy of a waiter in a restaurant to describe an API. When you dine out, you do not enter the kitchen to prepare your meal. Instead, you interact with a server who takes your order, communicates it to the kitchen, and returns your prepared food to you. In the digital world, APIs play a similar role. They act as intermediaries that allow one program to request information or services from another program without needing to understand the complex internal workings of that program. A real-world example of API usage is a weather app on your smartphone. The app itself does not collect or generate weather data. Instead, the app uses an API to send a request to a weather service's servers, receive current weather information, and display it in a user-friendly format.

For social work researchers, APIs give access to a wide range of AI services, data repositories, and automations, particularly suited to addressing complex social issues and analyzing diverse datasets. Consider a scenario where you have a collection of text documents in various languages that you need to have translated into English. Carrying out this task with an LLM is straightforward when you have a small number of documents. This involves copying and pasting the text into the web interface along with an instructional prompt requesting the LLM to perform a translation. However, this workflow becomes impractical when processing hundreds, thousands, or even millions of documents. Yet, with an API, this is a trivial task. You can establish a direct connection between your computer and the API service the LLM offers (e.g., OpenAI). You can then send your entire collection of documents programmatically for translation, saving significant time and effort. *Programmatically* refers to the process of automating tasks by writing code or using a software program, as opposed to doing them manually through approaches like copying and pasting.

**API Examples**

The APIs offered by the U.S. Census Bureau are an example of this technology that is relevant to social work researchers. Their geocoding API, for instance, allows researchers to convert lists of addresses into latitude and longitude coordinates, facilitating spatial analysis (U.S. Census Bureau, n.d.). Additionally, they offer a different API for programmatically downloading Census data from specified geographic areas. Downloading data from an API can be significantly more efficient than manually downloading multiple files and constructing datasets, especially for large-scale studies or when specific data subsets are needed.

The National Library of Medicine also offers a comprehensive set of API services through its Entrez Programming Utilities (National Center for Biotechnology Information, n.d.). These APIs provide programmatic access to various databases, with PubMed and PMC (PubMed Central) particularly useful for social work researchers. Unlike the standard web interface, these APIs allow researchers to automate the retrieval of article metadata and full-text content (where available). This capability is especially valuable for conducting different literature reviews, including systematic reviews, meta-analyses, and scoping reviews. These APIs also provide access to the complete collection of open-access journal articles, which can be the basis for automatically retrieving and disseminating articles in community organizations that do not have access to subscription-based journals.

**Features of Web-Based APIs**

The primary purpose of an API is to facilitate communication between software programs either locally on a single computer or network or externally, linking a local software program to a server housed elsewhere. For example, Perron and colleagues (2024) used an API to connect Python with *local* LLMs to securely perform a classification task on a collection of sensitive, unstructured child welfare documents. However, this article focuses on external, web-based API services. These external APIs are particularly valuable for social work research because they connect local systems to remote services and databases, offering capabilities beyond what a single computer or local network can achieve.

At the heart of these web-based API services are endpoints. An endpoint is a specific URL that serves as the point of contact between your application and the API service. It's essentially an address where your application sends requests to access particular functions or data from the API. For example, the endpoint for accessing ChatGPT models from OpenAI is [https://api.openai.com/v1/models](https://api.openai.com/v1/models) (OpenAI, n.d.). This URL becomes an integral part of the code interacting with the API.

While many data processing tasks can be performed using freely available software packages run locally on the user's computer, API service providers often offer more efficient and sophisticated solutions that require minimal effort from the researcher. For instance, free Python packages are available when analyzing unstructured text documents saved in PDF format but often require researchers to write and maintain code for data conversion. In contrast, specialized API services can perform this task more efficiently and accurately through a simple request to the appropriate endpoint.

**Types of Web-Based APIs**

The API service ecosystem is rapidly evolving, with new providers emerging and existing services expanding their capabilities regularly. For this reason, we'll discuss broad categories of APIs and their potential applications in social work research, focusing on the types of endpoints they offer:

1. Natural Language Processing (NLP) APIs: Endpoints for analyzing textual data, allowing researchers to process large volumes of unstructured text, extract key themes, sentiments, and entities from sources such as client narratives, policy documents, or social media content.

2. Social Media APIs: Endpoints for accessing data from social media platforms, allowing historical and real-time data streams for tracking public sentiment, identifying emerging social issues, and analyzing social networks.

3. Government and Public Data APIs: Endpoints for accessing official statistics, datasets, and metadata from governmental agencies at the federal, state and local levels.
4. Geospatial and Mapping APIs: Endpoints to analyze and visualize data in a geographical context, revealing patterns and disparities across different locations.
5. Survey and Data Collection APIs: Endpoints for gathering primary data and streamlining survey creation, distribution, and analysis.
6. Health and Wellness APIs: Endpoints to access health data from a fitness tracker, health monitor, or another device that captures biometrics and physical activity.
7. Web Search and Scraping APIs: Endpoints enabling automated web searching and scraping.
8. Other Machine Learning and AI APIs: Endpoints offering AI capabilities for uncovering complex patterns and relationships within data.

Social work researchers can find these API services and their specific endpoints through strategic internet searches using keywords related to their research needs. LLMs like ChatGPT or Claude can also be valuable resources in this process. Researchers can leverage LLMs to find information about relevant APIs and get assistance in writing the initial code to interact with these endpoints. For instance, a researcher could ask an LLM for suggestions on APIs suitable for analyzing social media sentiment in the context of community health studies. The LLM might provide information about relevant APIs, including their endpoint URLs, and even offer sample code snippets for making API calls.

**Cost Structures, Dashboards, and Playgrounds**

Understanding the cost structures, management tools, and testing environments for APIs is essential before integrating them into a research workflow. Many API providers offer a tiered pricing approach, often starting with a free tier to encourage adoption and exploration. Free tiers typically come with usage limitations, the most common being rate-limiting. This restricts the number of API requests within a specific timeframe, analogous to how public utilities prevent individual users from monopolizing resources. For instance, an API might allow 1,000 requests per hour or 10 per second. Exceeding these limits usually results in temporary blocks or error responses.

To manage these aspects, API providers often offer dashboards - web-based interfaces that serve as control centers for API usage. These dashboards provide real-time insights into usage metrics, helping researchers track their consumption against rate limits. They also often include cost calculators, allowing researchers to estimate expenses for accessing higher-paid tiers based on projected usage. For those considering paid API services, dashboards become even more critical. They offer detailed breakdowns of API calls, data transfer volumes, and associated costs. This transparency helps researchers make informed decisions about resource allocation and budget management. Paid tiers typically offer higher rate limits, priority support, and access to premium features or datasets, making them suitable for large-scale or ongoing research projects.

Pricing models for paid APIs vary. Subscription-based plans charge fixed recurring fees for a set access or usage allowance level. Pay-per-use models, conversely, bill based on actual consumption. In this model, users are charged a small fee for each request they make to the API, such as $0.01 per API call, allowing for predictable costs based on usage. Other providers offer hybrid models, combining a base subscription with overage charges for excess usage. Dashboards play a vital role here, often including features like usage alerts to prevent unexpected costs.

Another valuable tool in the API ecosystem is the playground or sandbox environment. These interactive interfaces allow researchers to experiment with API calls, test different parameters, and see immediate results. Playgrounds are invaluable for understanding an API's capabilities and limitations before full integration into a research workflow. They often come with code generation features, producing ready-to-use snippets in various programming languages based on the tested API calls.

**Data Security, Privacy Considerations, and Ethical Concerns**

While APIs can offer powerful capabilities that enhance research efficiency and scope, they should be used judiciously and with a clear understanding of the potential risks to data security and privacy. Sending data to off-site providers introduces potential risks that researchers must carefully consider and mitigate. This is particularly important when dealing with sensitive information or data subject to confidentiality requirements of the Institutional Review Board (IRB).

When you request an API, the data you send leaves your local system and is processed on the service provider's servers. This means you may not have complete control or visibility into how this data is handled, stored, or used beyond your immediate request. Data transmission and external processing could violate IRB confidentiality requirements if not adequately managed. To address these concerns, researchers should take several precautionary steps. First and foremost, carefully review the API provider's privacy policy and data handling practices. Ensure that the API uses secure transmission protocols to protect data in transit. Whenever possible, anonymize or de-identify data before sending it through an API to minimize the risk of exposing sensitive information. Consider prioritizing API services that offer end-to-end encryption for an additional layer of security, and be cautious about sending personally identifiable information or other sensitive data through external APIs. Verifying that the API provider complies with relevant data protection regulations is also essential. These precautions will likely mitigate many of the risks associated with API use, but will not resolve data privacy and security issues in all instances. When working with sensitive data, local software can be used as an alternative to cloud-based providers (see Perron et al., 2024).

The social work literature on AI includes extensive discussions about ethical concerns, including potential bias in analysis and outputs (Goldkind, 2021; Patton et al., 2023; Reamer, 2023; Victor et al., 2023; Victor et al., 2024). Model bias in AI refers to systematic errors in AI outputs that can lead to unfair or discriminatory outcomes. These biases often stem from the data used to train the models, the algorithms employed, or the design choices made during model development. These concerns are equally relevant when utilizing AI models through APIs. However, it's necessary to understand that APIs do not directly contribute to model bias. Instead, APIs are an alternative method of connecting to AI models and services, offering programmatic access rather than interaction through a web interface. When accessing AI models through APIs, researchers are interacting with the same underlying models that would be accessed through web interfaces. Therefore, any biases inherent in these models remain present, regardless of the access method.

**Leveraging LLMs through APIs**

So far, we have focused broadly on APIs. In this section, we focus on understanding how to leverage the power of LLMs through an API. This method has numerous advantages over copying and pasting information into the web interface. The most significant advantage of using an API is scalability. APIs enable efficient processing of large volumes of data, allowing researchers to automate tasks that would be impractical through a web interface. For instance, a researcher could use an API to link their data to an LLM and automatically translate hundreds or even thousands of article abstracts from one language to another, rather than manually entering each into the web interface, significantly expediting the literature review process for international studies.

APIs also give researchers enhanced control over LLM parameters and specifications (such as temperature, sampling methods, and response constraints). This fine-grained control of adjustable parameters allows for more precise and tailored use of LLMs, making them better suited to specific social work research needs. A key example is the ability to adjust the *temperature* parameter, influencing the creativity and randomness of the LLM's outputs. Lower temperatures produce more predictable, focused responses - ideal for tasks requiring factual accuracy. Higher temperatures result in more diverse and creative outputs, which might benefit idea generation. Furthermore, APIs offer greater flexibility in output customization. Researchers can specify the desired length and format (e.g., JSON, XML, Markdown) of LLM responses, allowing for precise tailoring to research needs. This feature is handy when extracting and structuring information from unstructured text documents, a common task in qualitative social work research.

Many LLM APIs also allow for system-level instructions or context settings. This capability enables researchers to 'prime' the LLM with relevant background information, methodological approaches, or ethical considerations specific to their social work research. By setting the appropriate context, researchers can significantly improve the relevance and accuracy of the LLMs' outputs for specialized research tasks.

**Examples of Using APIs with LLMs**

To demonstrate the applicability of these technologies to social work, we highlight two recent studies by Luan et al. (20024a, 20024b) that show how API-facilitated use of LLMs can enhance existing research methodologies in our field. Luan and colleagues conducted a scientometric review of research on left-behind children in mainland China and a social network analysis of scholars publishing on this topic. They used an API to interact with ChatGPT, efficiently processing article metadata from over 1,200 articles. This method showcases the power of combining API access with LLM capabilities to handle large-scale data analysis tasks.

One particularly illustrative example is extracting structured information from unstructured text. The researchers needed to determine the geographic location of each study, information typically contained in article abstracts but in an unstructured format. Manually reviewing each abstract to record the province and city was unfeasible, especially given the bilingual nature of the articles (English and Chinese) and the limited number of bilingual team members. Luan and colleagues leveraged ChatGPT

13through an API to extract and structure the information to overcome this challenge. The LLM accurately processed this data in both languages. A manual review of 100 abstracts in Chinese and 100 in English showed perfect accuracy, confirming that this task could be reliably offloaded to the LLM in the research workflow.

We highlighted this particular task because it exemplifies a widely applicable approach in social work research. Researchers encounter similar challenges when processing unstructured text data from various sources, including survey responses, interview transcripts, and bibliometric information. This demonstration also showcases the multilingual capabilities of LLMs. Significantly, the API code used for this text extraction can be readily adapted for other research-relevant functions, such as text translation, summarization, or classification, making it a versatile tool for social work researchers.

**Step-by-Step Guide to API Use with LLMs**

To provide more detailed instructions on API use with LLMs, this section describes each step in the process, using the recent research by Luan et al. (2024a; 2024b) as a backdrop. We provide the individual snippets of code used to read the data, connect with ChatGPT through the API, and receive the final response. We then give a breakdown of each part of the script, including the whole script in Appendix A for the reader's convenience. This example was chosen because the extraction task is a generalized template that can be swapped with other natural language processing tasks (e.g., summarization, classification, title generation, keyword extraction). Moreover, the data source can be easily replaced with different text data, such as newspaper articles, open-ended survey responses, or social media posts.

As mentioned, Python has emerged as the primary programming language for interacting with LLMs. Python is freely available, and we use this programming language here for demonstration (Python Software Foundation, 2023). After installing Python, the user will need an Integrated

Development Environment (IDE), a software application for writing code and interacting with Python. Many IDE options are available, with Visual Studio Code (Microsoft Corporation, 2023) being popular. Another option is to use an online Python environment, such as Google Colaboratory, often called Google Colab ([colab.research.google.com](colab.research.google.com)). Google Colab is a free, cloud-based service provided by Google that allows you to write and run Python code directly in your web browser. It requires no installation or configuration on your local computer, making it ideal for researchers who may not have prior programming experience or access to specialized software. This demonstration assumes that Python is installed on a local computer rather than a cloud-based environment.

We further note that the code is specified for a Windows operating system. Slight differences are required when working with macOS. The code we provide here helps show the logic, structure, and procedures. The reader is encouraged to refer directly to any service provider's most recent documentation, as small code setup changes often occur over time.

**Obtaining and Storing an API key**

Before interacting with an API using Python, you must obtain an API key from your chosen service provider. To acquire this key, perform an internet search for "API key" along with the name of your service provider. Follow the provider's instructions to set up an account and obtain the key, noting that many API services require payment. We use OpenAI as the provider for this demonstration, but the process is similar to other services such as Anthropic or Google. The API key is typically a long sequence of numbers and letters.

Once you have your API key, it should be stored securely. Create a file named ".env" in your project directory. To do this, open any text editor like Notepad (on a PC) TextEdit (on macOS), and save a new file to your directory with the name .env (including the dot). This special file type is used for safe and secure storage of sensitive information. In the .env file, add your API key in the following format:

OPENAI_API_KEY=your_actual_api_key_here

For example, if you were using OpenAI, your .env file might contain a line that looks like this (using a fictitious key):

OPENAI_API_KEY=sk-proj-92C5jX0_ofUYaMoahk6I5azdAmr-oTKa4PvSyIz9rPJ32T

By storing your API key in this manner, you keep it separate from your main code, enhancing security and making it easier to manage across different projects or environments.

**Setting up the Environment and Data**

As the next step, we must create a Python environment to write and execute the necessary code. This setup is specific to Python programming and is separate from the API. The following code involves importing necessary libraries that provide the tools we need to perform the extraction task.

```python
import pandas as pd
from openai import OpenAI
import os
from dotenv import load_dotenv
os.chdir(r"USER_DIRECTORY_PATH_HERE")
df = pd.read_excel(r"LBC_MetaData.xlsx")
```

These lines of code set the stage for our data processing and API interaction. We begin by importing the pandas package into the environment, giving it an abbreviation of pd for convenience. The pandas library provides a robust toolset for handling data. We import the OpenAI function from the openai library. This function is used to exchange data and responses from ChatGPT through Python with the API. The os library offers functions for interacting with the operating system. We also import the load_dotenv function from the dotenv package. This provides a safe and secure way for loading our API key, which is explained in further detail in the following section.

We then use the os.chdir to specify the directory or folder on our computer where our files are saved. The heart of our data setup is in the last line: df = pd.read_excel(r"LBC_MetaData.xlsx")  Here, we're using pandas' read_excel() function to load our Excel file into Python. The r before the file path creates a raw string, which is necessary for reading a path when using a PC. A slightly different specification is necessary for macOS users. When we call read_excel(), pandas reads the Excel file and converts it into a data frame, which we assign to the variable df. Using pandas to read our Excel file, we've effectively bridged the gap between our raw data and the Python environment where we'll process it. This data frame (df) can be easily inspected, cleaned, and prepared before we send it to the API for processing. We can leverage additional various pandas functions to filter, transform, or analyze our data as needed.

**Authentication: Securing API Access**

Before using the API, we need to set up the service, also called API authentication. This critical security step is standard for most API services and is analogous to logging into a website with a username and password. The following lines of code are used to authenticate:

load_dotenv()

client = OpenAI()

      model = "gpt-4o"

The .env file constructed in the first step should be moved to the working directory set using os.chdir. The load_dotenv() function loads environment variables from a .env file in the current working directory. This file contains key-value pairs, including the API key. This method keeps the API key separate from the code, enhancing security and allowing easier management across different environments. Thus, you can easily share or display code online without exposing your API key. The .env file should be kept confidential, like a password.

      The client = OpenAI() is a function that reads the API key and also configures the API endpoint. The URL that defines the endpoint is automatically included in the OpenAI function in the client = OpenAI() code snippet. This tells Python where to send its API requests, like entering a web address in a browser that directs us to a specific website. The URL can be easily changed using a different AI provider following the OpenAI API request format. This standardization means switching between AI services by changing the model name, API key, and base URL according to each provider's documentation. The online GitHub repository for this project (https://github.com/beperron/demystifying-APIs) provides examples of these code modifications. This flexibility is particularly valuable in research settings, where different models may perform better for specific tasks or offer more cost-effective solutions. Researchers can quickly adapt their code to work with various providers, allowing rapid iteration while effectively managing resource constraints.

**Defining the Extraction Prompt**

      The next step involves defining the extraction task by crafting a prompt for the LLM that serves as specific instructions. While this approach is similar to the web interface, where we provide the LLM with instructions (a prompt) and the abstract, our API-based workflow differs as we're processing approximately 1,200 abstracts. In the API context, we define two separate variables: the system and the

user prompt. The system prompt sets the context for the LLM's role, while the user prompt contains the specific task instructions. For this demonstration, we use a standard system prompt, though it can be adapted for more specific roles:

> system_extraction_prompt = "You are a helpful assistant with expertise in social work research and Chinese geography."

The user prompt provides detailed instructions for the extraction task. We define this prompt in Python using triple quotes, allowing it to span multiple lines for better readability and organization.

> user_extraction_prompt = """ Carefully read the following scientific abstract. Identify and extract any Chinese locations that are first-level administrative divisions under China's central government in mainland China. These include:
>
> - Provinces
>
> - Municipalities directly under the central government
>
> - Autonomous regions
>
> If a city or county is mentioned instead of a province or autonomous region, return the corresponding province, municipality, or autonomous region in English. For example, if '广州市' (Guangzhou) is mentioned, return 'Guangdong Province'. Only extract locations at this administrative level. If multiple locations are present in a single abstract, separate each location with a semicolon.
>
> Return NONE if no specific province, city, or administrative region is mentioned. For unspecified mentions (e.g., "10 provinces", "several regions", "southwestern China", "western China"), return 'NONE'. Do not make any assumptions or inferences about unspecified locations. Return NONE if no specific name is mentioned, or need to guess or infer any locations based on context or general knowledge."""

This two-part prompt structure allows for precise control over the LLM's behavior, ensuring accurate and consistent information extraction across the large dataset of abstracts.

**Defining the Extraction Function**

The next step involves creating an extraction function applicable to each data row. This function incorporates the system prompt, user prompt, and additional LLM parameters to set up an API request. API interactions send a request (or call) from our local computer to the remote server (e.g., OpenAI) with specific instructions and data. For our example, we develop a specialized function for extracting the location information from each abstract. This function handles the extraction logic and defines API service instructions, including model selection and other detailed controls unavailable through the web interface. These controls enable precise customization of AI behavior and output to meet specific research requirements. The following code block presents the complete data extraction function. While this may seem complex to readers unfamiliar with Python, we will provide a step-by-step breakdown to explain each component and its role. We also want to reemphasize that LLMs excel at explaining code, and encourage readers to use LLMs to clarify any challenging parts of the code, making the learning process more accessible, even for programming novices.

```python
def extract_location(abstract):
    response = client.chat.completions.create(
        model=model,  # Specify the model to use
        messages=[
            {
                "role": "system",
                "content": system_extraction_prompt
            },
```

```
      {
        "role": "user",
        "content": user_extraction_prompt + abstract
      }
    ],
    temperature=0,
    max_tokens=2500,
    top_p=1,
    frequency_penalty=0,
    presence_penalty=0
  )
  location = response.choices[0].message.content.strip()
  return location
```

To help clarify the code above, we'll describe each function component. The def extract_location(abstract): line defines our function name and specifies that each abstract in the dataset is the input text. We emphasize that this function processes each abstract separately rather than sending the entire file to the LLM. The client.chat.completions.create method is used to send a request to OpenAI for each abstract. In the preceding code block, we already specified our model, which is then passed to the function in the model = model. The messages parameter defines the interaction with the AI assistant, including the system and user messages we previously discussed. The + abstract at the end of the user content inserts the raw abstract data directly into the prompt, tailoring the analysis to our specific research data. This part of the code can be easily adapted to perform different tasks or outputs, such as extracting the author's university instead of the location.

The parameters temperature=0, max_tokens=2500, top_p=1, frequency_penalty=0, and presence_penalty=0 control the characteristics of the AI's response. These optional parameters provide additional control over data processing, which is unavailable in the web interface for cloud-based LLMs. Temperature, top_p, and the frequency penalty influence the model's creativity, while the tokens specification limits the model's text output length.

The line location = response.choices[0].message.content.strip() is crucial because the API response contains more information than the extracted location names. It targets the text output generated by the AI model within the complex response object. The .strip() function removes any extraneous whitespace from the AI's response. This process ensures we're isolating only the relevant information - the list of locations - from the broader response object, preparing the extracted data for further processing or analysis in our program.

**Row-by-Row API Processing**

The final step of our procedure is the actual data processing method, known as *batch* processing, which facilitates the automation of analyses with the API. This approach is necessary when dealing with large volumes of data. We set up the function in the earlier step, which we then apply to each row of our data. Essentially, we're creating a digital assembly line for our data, where each piece is handled individually, processed, and then stored before moving on to the next. The core of this process is encapsulated in a single line of Python code:

df["Location"] = df["Abstract"].apply(extract_location)

This line represents a sophisticated form of data manipulation. Here, 'df' refers to our DataFrame, a two-dimensional data structure in Python that holds our research data, similar to a spreadsheet. The apply function, a powerful feature of the pandas library in Python, is the key to our batch-processing approach. It systematically applies our custom function `extract_location` to each entry in the `Abstract` column of our data.

The apply function takes each abstract and processes it with the `extract_location` function. The function uses the API to send the abstract, prompt, and other model specifications to ChatGPT as part of this process. ChatGPT carries out the extraction task and returns the response via the API. The new response is stored in a new column named `Location`. This process repeats for each abstract in our raw data until all data have been processed. The final line of our code writes a new Excel file containing the extracted data: `df.to_excel("LBC_Locations_Extracted.xlsx")`.

This approach offers several advantages for social work research. It automates what would otherwise be a tedious and error-prone manual process, saving time and reducing the potential for human error. It also allows for consistent data processing, ensuring that each piece of information is handled the same way, which is necessary to maintain the integrity of research findings. Moreover, this method is highly scalable. Whether working with a small pilot study or a large-scale research project, the code remains the same; only the processing time changes. This scalability makes it an invaluable tool for longitudinal studies or research projects that may grow in scope over time.

**Constructing API Calls: A Practical Approach for Social Work Researchers**

While APIs offer powerful research capabilities, implementing them can be challenging for researchers without programming backgrounds. This section provides a practical framework for API integration, showing how LLMs can support the process. We recognize that the technical nature of API implementation may be unfamiliar to some readers. For those new to programming concepts, we recommend first reviewing the glossary of terms in Appendix A and considering the following approaches: reading through the conceptual overview before diving into the code examples, consulting with technical colleagues, or using LLMs like ChatGPT or Claude to break down complex concepts.

While this section contains detailed code examples, understanding the core concepts is more important than mastering the technical specifics. We present technical code examples to demonstrate API call structure and implementation. For reference, all code samples are available in this paper's Appendix and our GitHub repository: (https://github.com/REDACTED/demystifying-APIs). Readers can revisit these examples as their comfort with API concepts grows.

We focus on two key aspects of LLMs: foundation knowledge and in-context learning. Foundation knowledge refers to information embedded in LLMs during their training, which may include API documentation. Researchers can access this knowledge through strategic prompting, comparing the model's output with official API documentation. However, verifying and testing any generated code is essential, as LLMs may occasionally produce inaccurate information. The second approach, in-context learning, involves providing the LLM with specific segments of API documentation to assist in formulating API code. This is particularly useful for newer APIs or when seeking up-to-date information. Both strategies require effective prompting to generate accurate and helpful responses.

Our method guides researchers through understanding API documentation, crafting initial API calls, testing and refining these calls, scaling up to handle larger datasets, and implementing error handling and optimization. Throughout this process, LLMs assist at each stage, enabling social work researchers to gradually build expertise in API usage. This method simplifies technical aspects of API usage, allowing researchers to focus on their core competencies and research questions. To illustrate the proposed method, we assume a large-scale analysis of service providers' geographic distribution. We have individual addresses for each provider but need latitude and longitude coordinates to facilitate the analysis.

**Locating and Understanding the API**

Our search for a suitable geocoding API blended traditional research methods with modern AI technologies. We conducted thorough online searches, which provided an overview of available geocoding API options. To complement this, we used LLMs like ChatGPT and Claude. These systems, trained on vast datasets, offered insights based on their foundation knowledge. Through this process, the U.S. Census Bureau's Geocoding API emerged as a standout option due to its free accessibility and the fact that it does not require an API key. These attributes made it ideal for our working example, lowering the barrier to entry and simplifying implementation for potential users.

After identifying the U.S. Census Bureau's Geocoding API as our chosen service, we focused on understanding its functionality and requirements by reviewing the official documentation. We used LLMs to simplify the technical jargon. To illustrate, we used the following prompt with different LLMs, providing the source documentation by copying and pasting the U.S. Census Bureau's webpage into the prompt:

> Here is the documentation for the U.S. Census Bureau's API for geocoding. Provide a plain language summary of this API, explaining what it does and its requirements. Your response should explain all technical concepts.

This process confirmed the API's suitability for our use case. Although the documentation contained technical terminology, LLMs provided clear explanations through repeated prompting.

**Crafting a Single API Call**

The next step was to generate code to make a single API call with a given address. We were not processing a large dataset but getting a minimal working example at this stage. This approach saves time and resources, notably if the API incurs costs, as testing on large datasets could result in unnecessary charges without properly processed data. For this example, we used in-context learning, providing the

LLM with specific documentation to help construct the API call. We used the following prompt with ChatGPT-4:

> Here is the online documentation for the U.S. Census Bureau API. Help me generate Python code to obtain the latitude and longitude coordinates for a single address: 1080 South University Avenue, Ann Arbor, MI 48109. Provide me with the code to run this in a Python notebook. [Insert documentation here.]

We also tested the LLM's capability by asking it to generate the code based on its foundation knowledge:

> Using the U.S. Census Bureau API, help me generate Python code to obtain the latitude and longitude coordinates for a single address: 1080 South University Avenue, Ann Arbor, MI 48109. Provide me with the code to run this in a Python notebook.

The code was tested in a Google Colaboratory notebook, an environment for running Python code noted previously. The code generated from both prompts was processed without error and returned the expected latitude and longitude coordinates. If an error had been encountered, we would have provided the LLM with the error message for further refinement. This underscores the importance of proficiency in prompting LLMs which is often iterative in nature. The code produced by ChatGPT-4 is provided in Appendix B.

**Scaling Up API calls**

With a working example, we scaled up our code to process many addresses. Assume we have an Excel spreadsheet (Addresses.xlsx) with a column called "Address." Each row contains an individual address for a service provider. Using our functioning prompt for a single address, we asked the LLM to generate code to process the data file. We included the entire code for processing a single address as context, tasking the LLM with generalizing a functioning example. The following prompt was used:

> Here is the code for obtaining latitude and longitude coordinates for a single address using the U.S. Census Bureau's API. Adapt this code to process addresses in an Excel file called Addresses.xlsx. The addresses are in the 'Address' column, and the results should be saved in new columns labeled 'Latitude' and 'Longitude.' Provide comments in the code explaining each step.

The adapted code produced by ChatGPT-4 is included in Appendix C.

## Conclusion and Next Steps

This article provides a foundational understanding of how APIs can unlock the capabilities of LLMs and other AI-driven web services to enhance social work research. The traditional focus in social work has been mainly on statistical programming, yet current data demands necessitate general programming skills that involve developing and integrating coding-based solutions, including APIs. APIs are crucial for expanding data capabilities in a highly scalable manner, whether interacting with external web-based services or local language models when security needs are required. Their role goes beyond accessing remote services; they also provide the infrastructure needed to automate processes and create more efficient workflows.

While this article primarily discusses coding-based API solutions, it is important to note the availability of no-code tools for API interactions. These tools, although user-friendly, often lack the customization and advanced functionality required for specialized research needs. Researchers looking to implement unique and tailored solutions will likely need to engage with code to overcome the limitations of no-code interfaces. Our emphasis on coding-based approaches is intended to foster a deeper understanding of these tools, enabling social work researchers to fully leverage the power of APIs for both specialized tasks and broad-scale analyses.

APIs also hold immense promise for enhancing social work education and practice. Learning management systems like Canvas provide API services to help instructors manage course data efficiently, facilitating more personalized and automated student engagement. APIs are also the building blocks of sophisticated workflows that can automate routine tasks, allowing social workers to focus on more impactful aspects of their work. Integrating these technologies into practice and education can modernize the field, making data more accessible and processes more efficient.

The ability to interact effectively with APIs is not just an advanced technical skill—it represents a significant shift in how data is approached in social work research, practice, and education. For social workers to remain relevant and make impactful contributions in an increasingly data-driven world, embracing these tools is not optional but necessary. APIs provide the scalability, efficiency, and flexibility required to advance the scope and impact of social work research, ultimately contributing to improved social services and better outcomes for communities.

# Appendix A: API Request to Extract Country Names From Unstructured Text

The following Python code demonstrates extracting the country name from unstructured text in article meta-data. All hashtags are comments within the code that Python does not process.

```python
# Setup workspace
import pandas as pd
from openai import OpenAI
from dotenv import load_dotenv
import os
# Load the Excel file into a DataFrame
os.chdir(r"USER_DIRECTORY_PATH_HERE")
df = pd.read_excel(r"LBC_MetaData.xlsx")
# Setting up the Environment and Data
load_dotenv()
model = "gpt-4o"
client = OpenAI()
# Define extraction prompts
system_extraction_prompt = "You are a helpful assistant with expertise in social work research."
user_extraction_prompt = """Carefully read the following scientific abstract. You must identify any
    Chinese province, city, and special administrative region noted in the abstract. If multiple
    locations are present in a single abstract, separate each location with a semicolon. Return NONE
    if no specific province, city, or administrative region is mentioned. You must perform this task
    accurately. Do not make up or guess any information."""
# Define extraction function
```

```python
def extract_location(abstract):
    response = client.chat.completions.create(
        model="gpt-4o",  # Specify the model to use
        messages=[
            {
                "role": "system",
                "content": system_extraction_prompt
            },
            {
                "role": "user",
                "content": user_extraction_prompt + abstract
            }
        ],
        temperature=0,
        max_tokens=2500,
        top_p=1,
        frequency_penalty=0,
        presence_penalty=0
    )
    location = response.choices[0].message.content.strip()
    return location

# Row-by-row processing
df["Location"] = df["Abstract"].apply(extract_location)
```

```python
df.to_excel("LBC_Locations_Extracted.xlsx")
```

# Appendix B: API Request to Geocode a Single Address

ChatGPT-4o generated the following Python code. All hashtags are comments within the code that Python does not process.

```python
import requests
# Define the API endpoint and parameters
url = "https://geocoding.geo.census.gov/geocoder/locations/onelineaddress"
params = {
    'address': "1080 South University Avenue, Ann Arbor, MI 48103",
    'benchmark': 'Public_AR_Current',
    'format': 'json'
}
# Make the request to the API
response = requests.get(url, params=params)
# Parse the response
data = response.json()
# Extract latitude and longitude
coordinates = data['result']['addressMatches'][0]['coordinates']
latitude = coordinates['y']
longitude = coordinates['x']
# Output latitude and longitude
latitude, longitude
```

**Appendix C: API Request to Geocode a Collection of Addresses**

ChatGPT-4o generated the following Python code. All hashtags are comments within the code that Python does not process.

```python
import requests

import pandas as pd

# Read the Excel file and load the addresses into a DataFrame

df = pd.read_excel('Addresses.xlsx')

# Initialize lists to store latitude and longitude results

latitudes = []

longitudes = []

# Loop through each address in the DataFrame

for address in df['Addresses']:

    # Define the API endpoint and parameters

    url = "https://geocoding.geo.census.gov/geocoder/locations/onelineaddress"

    params = {

        'address': address,

        'benchmark': 'Public_AR_Current',

        'format': 'json'

    }

    # Make the request to the API

    response = requests.get(url, params=params)

    # Parse the response

    data = response.json()
```

```python
    # Extract latitude and longitude, if the address is matched
 if data['result']['addressMatches']:
        coordinates = data['result']['addressMatches'][0]['coordinates']
        latitudes.append(coordinates['y'])
        longitudes.append(coordinates['x'])
    else:
        # If no match is found, append None values
        latitudes.append(None)
        longitudes.append(None)
# Add the latitude and longitude lists as new columns to the DataFrame
df['Latitude'] = latitudes
df['Longitude'] = longitudes
# Save the updated DataFrame to a new Excel file
df.to_excel('Addresses_with_Coordinates.xlsx', index=False)
```